\documentclass[twocolumn,superscriptaddress,aps,prl]{revtex4-1}

\usepackage{ulem,bm}
\usepackage{amsmath, upgreek, amssymb, graphicx, paralist, gensymb,epstopdf}
\usepackage[colorlinks, linkcolor=blue, citecolor=blue, urlcolor=blue]{hyperref}
\usepackage{array,epsfig,amsmath,amssymb}
\usepackage{graphicx}
\usepackage{dsfont}
\usepackage{color}
\newcommand{\ket}[1]{|#1\rangle}
\newcommand{\bra}[1]{\langle#1|}

\begin{document}

%\title{Quantum teleportation between multiple senders and receivers}
%\title{Quantum teleportation of shard quantum states}
%\title{Distributed quantum teleportation}
%\title{Distributed quantum teleportation with linear optics}
%\title{Distributed quantum teleportation of shared quantum information}
\title{Quantum teleportation of shared quantum secret}

\author{Sang Min Lee$^*$}
\affiliation{Korea Research Institute of Standards and Science, Daejeon 34113, South Korea}
\email{These authors contributed equally to this work.}
\author{Seung-Woo Lee$^*$}
\affiliation{Quantum Universe Center, Korea Institute for Advanced Study, Seoul 02455, South Korea}
\author{Hyunseok Jeong}
\affiliation{Center for Macroscopic Quantum Control, Department of Physics and Astronomy, Seoul National University, Seoul 08826, South Korea}
\author{Hee Su Park}
\affiliation{Korea Research Institute of Standards and Science, Daejeon 34113, South Korea}

\maketitle

%%%%%%% Introductory paragraph %%%%%%%%%%%%%%%%%%%%%%%%%%%%%%%%%%%%%%%%%

{\bf Quantum teleportation allows to transfer unknown quantum states between distant parties \cite{teleportation}. It is not only a primitive of quantum communications but also an essential task in realization of the quantum networks for promising applications such as quantum cryptography \cite{BB84,EK91} and distributed quantum computation \cite{Kimble2008}. Despite recent substantial progresses in the realization of quantum communications \cite{AdvancedTele}, teleportation of shared quantum information between multiple senders and receivers is still missing. Here we propose and experimentally demonstrate quantum teleportation among spatially separate parties in a quantum network. The protocol can be jointly performed by distributed participants, while none of them can fully access the transferred information. It can be generally extended to quantum communications incorporating an arbitrary number of parties without a trusted central node. Our work opens a route to the realization of distributed quantum communications and computations in quantum networks.}

%%%%%%%%%%%%%%%%%%%%%%%%%%%%%%%%%%%%

While the original teleportation protocol transfers quantum information from one place to another \cite{teleportation},  incorporation of multiple participants further merits consideration towards the realization of versatile quantum networks. Protocols to split quantum information from one sender to multiple receivers have been proposed~\cite{Karlsson98,Karlsson99,Hillery99} and demonstrated~\cite{opent}. With this protocol, no single receiver can fully access the information unless corporated by all the other receivers, constituting the basis of further extended quantum secret sharings \cite{QSS2,QSS3,TQSS1,GQSS1,GQSS2} or controlled teleportations \cite{Ctl1,Ctl2,Ctl4}. Teleportations of multi-party states have been studied \cite{ETT1,ETT2,ETT3}, however, a quantum teleportation between multiple senders and receivers has been missing so far. None of the previous protocols, to our knowledge, allow to transfer a shared or split quantum information among multiple parties directly to others without concentrating the information in any single location. The absence of such a protocol has thus led to the requirement of fully trusted central or intermediate nodes in the design of quantum communication networks \cite{Sun16,Pant17}.

In this letter, we propose and experimentally demonstrate a quantum teleportation between multiple senders and receivers. In this protocol, neither any single- nor sub-parties of senders and receivers can fully access the transferred quantum information. We report its experimental demonstration of connecting two senders and two receivers by entangled four photons. Our protocol relays quantum information over a network in a distributed manner without requiring fully trusted central or intermediate nodes. It can be further extended to include error corrections against photon losses, bit/phase-flip errors, and dishonest parties.

%\section{Protocol}

Suppose that a quantum secret in $\ket{{\cal S}}=\alpha\ket{0_L}+\beta\ket{1_L}$ with logical basis, $\ket{0_L}$ and $\ket{1_L}$, is shared by separated $n$ parties in quantum network, through a splitting protocol \cite{Karlsson98,Karlsson99,Hillery99}. We employ the GHZ-entanglement of photons to encode both the network and logical qubits (its extension to more general states are discussed later). The shared secret can then be written as $\ket{{\cal S}}=\alpha\ket{H}_{s_1}\cdots\ket{H}_{s_n} + \beta\ket{V}_{s_1}\cdots\ket{V}_{s_n}$, with horizontal $\ket{H}$ and vertical $\ket{V}$ polarizations of photons. The {\em senders}, i.e. a group of $n$ parties, attempts to transfer the secret to the {\em receivers}, i.e. another group of $m$ parties, connected in the network. None of the participants is fully trusted here so that no single or sub-parties of senders or receivers is allowed to access the secret during the whole process.

%{\it Distributed Bell-state measurement}: 
We introduce a {\em distributed Bell-state measurement} that can be jointly performed by separated parties. In general, Bell states of logical qubits with $n$ photons can be decomposed into combinations of $n$ two-photon Bell states (see Methods), and discriminated by performing $n$ times of standard Bell-state measurements ($\rm {\bf B}$). As its logical outcome is irrespective of the order of performed $\rm {\bf B}$ measurements, it is possible to separate all $\rm {\bf B}$ spatially or temporally with help of classical communications to share their results among the nodes where $\rm {\bf B}$ is respectively performed.

%%%%%%%%%%%%%%%%%%%%%%%%%%%%%%%%%%%%%%%
\begin{figure}[t]
\centering
\includegraphics[width=3.4in]{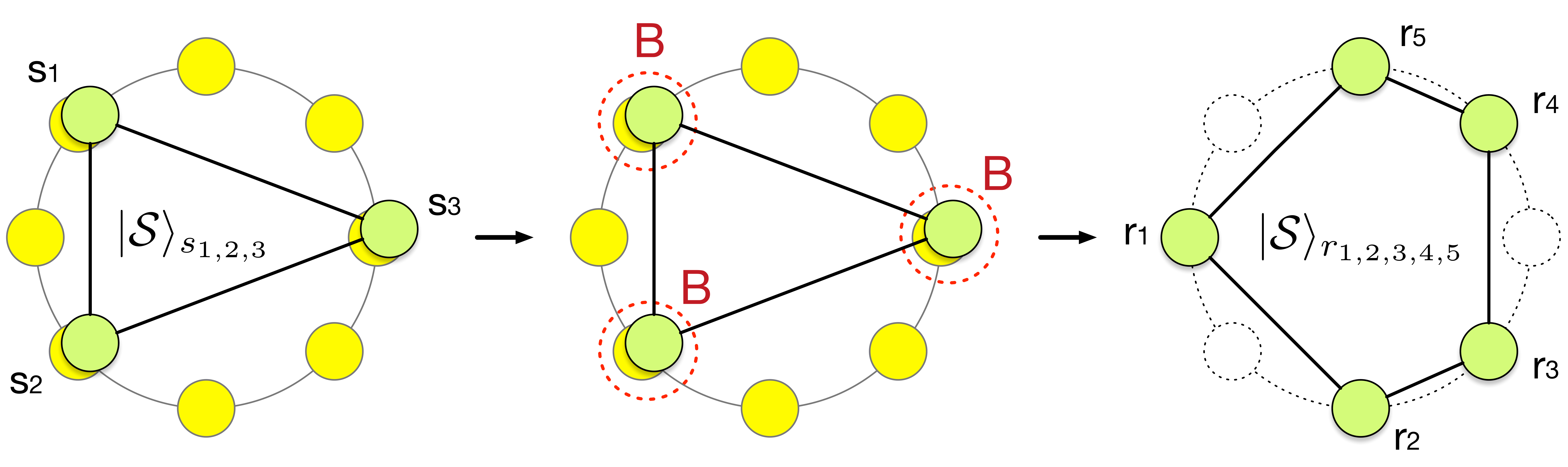}
\caption{Teleportation of quantum secret $\ket{{\cal S}}$ between multiple senders ($n=3$) and receivers ($m=5$) in a quantum network by distributed Bell-state measurements. After performing the Bell-state measurement $\rm {\bf B}$, each sender $s_i$ announces the result. Receivers can reconstruct and share the secret by appropriate joint work of local operations. No participants can access the secret during the procedures.}
\label{fig:tele}
\end{figure}
%%%%%%%%%%%%%%%%%%%%%%%%%%%%%%%%%%%%%%%%%%

%{\it Teleportation protocol}: 
The teleportation protocol between multiple parties in a quantum network is illustrated in Fig.~\ref{fig:tele}: Each separate sender performs $\rm {\bf {\bf B}}$ on two photons, one from $\ket{{\cal S}}_s$ and the other from the network channel, and announces the result. Conditioned on the results of all performed $\rm {\bf B}$, the reduced state of the channel at the receivers' locations is in $\ket{{\cal S}}_r=\alpha\ket{H}_{r_1}\cdots\ket{H}_{r_m} + \beta\ket{V}_{r_1}\cdots\ket{V}_{r_m}$ plus \textit{logical} Pauli operations \cite{SLee15}. The receivers jointly carry out appropriate \textit{local} Pauli operations according to the results announced by the senders to retrieve $\ket{{\cal S}}_r$.

For example, when $n=2$, teleportation of a shared secret $\ket{{\cal S}}_s= \alpha\ket{H}_{1}\ket{H}_{2}+\beta\ket{V}_{1}\ket{V}_{2}$ via network channel $(\ket{H}_{1'}\ket{H}_{2'})_s\ket{H}^{\otimes m}_r+(\ket{V}_{1'}\ket{V}_{2'})_s\ket{V}^{\otimes m}_r$ is explained by the joint state
\begin{equation}
\nonumber
\begin{aligned}
&(\ket{\phi^+}_{s_1}\ket{\phi^+}_{s_2}+\ket{\phi^-}_{s_1}\ket{\phi^-}_{s_2})(\alpha\ket{H}^{\otimes m}+\beta\ket{V}^{\otimes m})_r\\
+&(\ket{\phi^+}_{s_1}\ket{\phi^-}_{s_2}+\ket{\phi^-}_{s_1}\ket{\phi^+}_{s_2})(\alpha\ket{H}^{\otimes m}-\beta\ket{V}^{\otimes m})_r\\
+&(\ket{\psi^+}_{s_1}\ket{\psi^+}_{s_2}+\ket{\psi^-}_{s_1}\ket{\psi^-}_{s_2})(\alpha\ket{V}^{\otimes m}+\beta\ket{H}^{\otimes m})_r\\
+&(\ket{\psi^+}_{s_1}\ket{\psi^-}_{s_2}+\ket{\psi^-}_{s_1}\ket{\psi^+}_{s_2})(\alpha\ket{V}^{\otimes m}-\beta\ket{H}^{\otimes m})_r,
\end{aligned}
\end{equation}
where $\ket{\phi(\psi)^{\pm}}_{s_i}$ are the Bell state of the two photons in modes $(i,i')$ that sender $s_i$ holds. If the results of $\rm {\bf B}$s (which discriminate $\ket{\phi^-}$ and $\ket{\psi^-}$ out of four Bell states) of the two senders are $\ket{\phi^-}$ and {\it failure}, resectively, the case corresponds to the second term, therefore the receivers can reconstruct $\ket{{\cal S}}_r$ by a phase flip ($\hat\sigma_z$: $\ket{H}\rightarrow \ket{H}$, $\ket{V}\rightarrow -\ket{V}$) at any one receiver's location. Likewise for other results, the secret can be recovered by the receivers. It is straightforward to extend the protocol for arbitrary $n$ number of senders (see Methods). 

Any sub-parties cannot fully access the quantum secret during the teleportation procedures. For example, if one sender attempts to reconstruct the secret at his/her location based on the announced results by the other senders, the reduced state at his/her party is either $|\alpha|^2\ket{H,H}\bra{H,H}+|\beta|^2\ket{V,V}\bra{V,V}$ or $|\alpha|^2\ket{H,V}\bra{H,V}+|\beta|^2\ket{V,H}\bra{V,H}$ unless the whole channel is possessed by him/her (see Methods), therefore only the amplitude information is accessible. The same holds for any sub-parties of senders and receivers.

%\section{Experimental demonstration}

We demonstrate quantum teleportation between two senders ($n=2$) and two receivers ($m=2$) via a four-photon quantum network channel using total six photons.

\begin{figure*}[t]
\centering
\includegraphics[width=6 in]{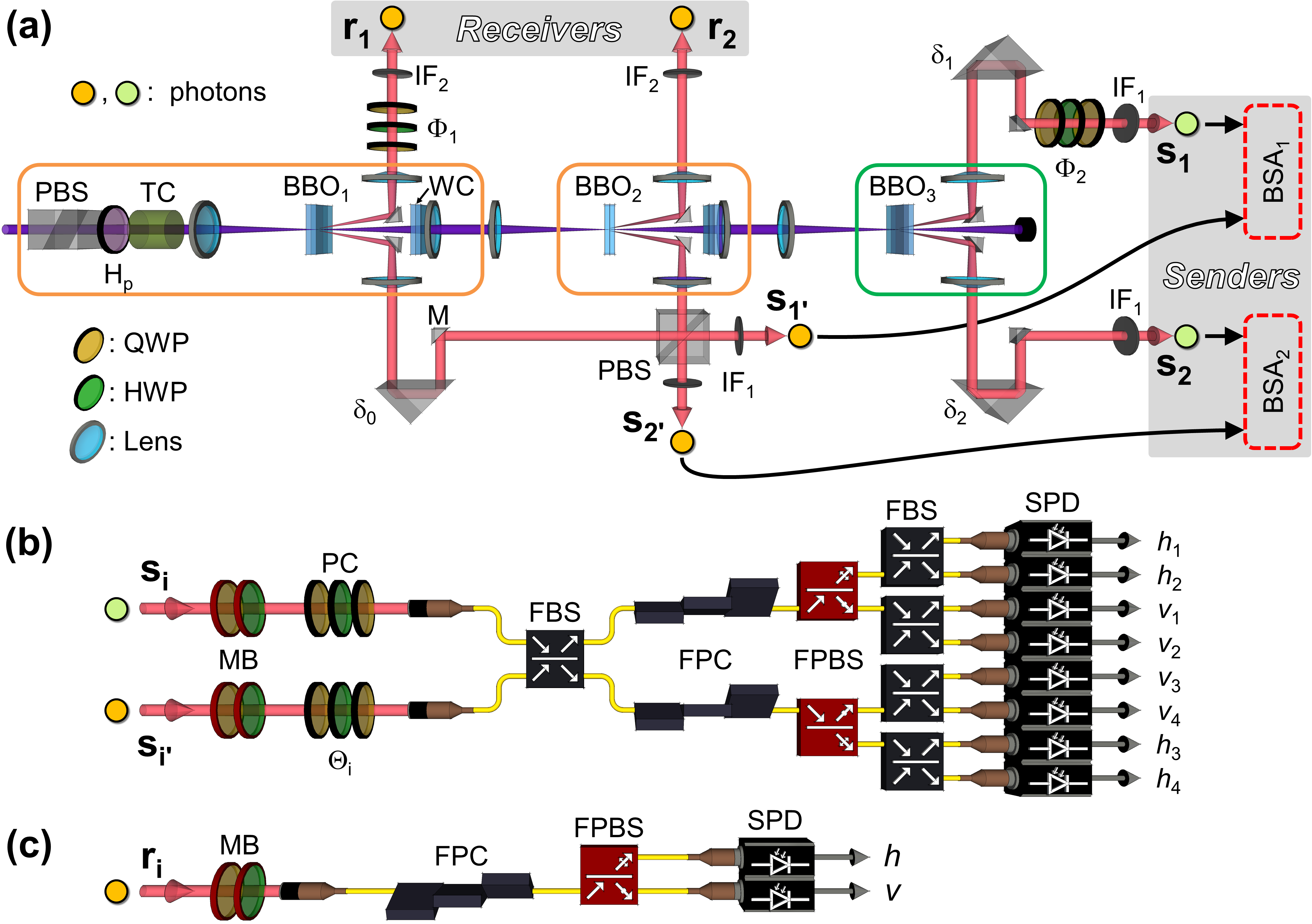}
\caption{Experimental setup. (a) Overall schematic for teleportation between two senders and two receivers. A four-photon GHZ state and a two-photon entangled input state are generated. (b) Structure of the optical-fiber-based local Bell state analyzer. (c) Polarization analyzer for single photons. M: mirror, PBS: polarizing beam-splitter, BBO: beta barium borate crystal pair (two mutually orthogonal 1-mm plates), TC: temporal walk-off compensator (16-mm-thick quartz plate), WC: spatial walk-off compensator (180$^\circ$-rotated BBO), QWP: quarter-wave plate, HWP: half-wave plate, H$_{\rm p}$: HWP for pump, IF$_i$: interference filter (half-maximum bandwidth of 3 nm and 20 nm for $i=1$ and $2$, respectively), MB: measurement basis controller, PC: polarization controller based on a combination of quarter-, half-, and quarter-wave plates (QHQ), FPC: fiber paddle polarization controller, $\Theta_i, \Phi_j$: relative phase shifter between the $H$ and $V$ polarizations using a QHQ, FBS: fiber beam splitter, FPBS: fiber polarizing beam splitter, SPD: single-photon detector.}
\label{fig:setup}
\end{figure*}
%{\it State generation and measurement}: 
Figure~\ref{fig:setup}(a) shows the schematic of our experimental setup. Photons are generated by spontaneous parametric down-conversion (SPDC) in BBO crystals (see Methods). Two polarization-entangled photon-pairs ($\ket{H}\ket{H}+e^{i \phi_k} \ket{V}\ket{V}$, $k=1,2$) generated by BBO$_1$ and BBO$_2$ are projected to a four-photon GHZ state, $\ket{\textrm{GHZ}_4}\equiv \ket{H}^{\otimes 4}+\ket{V}^{\otimes 4}$, by post-selection at modes $s_{1'}$ and $s_{2'}$~\cite{Zeilinger97}. The phase between the $\ket{H}^{\otimes 4}$ and $\ket{V}^{\otimes 4}$ components are set to zero by a phase shifter $\Phi_1$ in Fig.~\ref{fig:setup}(a), which is a combination of two quarter-wave plates (QWPs) whose slow axes are along 45$^\circ$ and one rotatable half-wave plate (HWP) in between. Input states of the form $\alpha\ket{H}\ket{H} + \beta\ket{V}\ket{V}$, where $\alpha$ and $\beta$ are complex constants, are generated through modes $s_1$ and $s_2$ by BBO$_3$ (green box). The magnitudes and the relative phases of $\alpha$ and $\beta$ are controlled by tilting the BBO crystals to change the coupling efficiency of the photons to the collecting single-mode fibers (SMFs) and by rotating the HWP in another phase shifter $\Phi_2$ at $s_1$, respectively.

Figure~\ref{fig:setup}(b) shows the structure of an optical-fiber-based Bell-state analyzer (BSA) that executes $\rm {\bf B}$. Optical fiber components such as fiber non-polarizing beam splitters (FBSs) and fiber polarizing beam splitters (FPBSs) replace the bulk optics components in the original design~\cite{Mattle96} to reduce the space and facilitate alignment. We note that two single-photon detectors (SPDs) are concatenated by an additional FBS at each output port of an FPBS. Coincidence counts (CCs) of the two SPDs identify the failure events of $\rm {\bf B}$. The conventional schemes using one SPD at each output port cannot discriminate these failure events from the errors caused by photon losses. The success probability of the failure detection is 50\% with the current scheme, and can reach near-unity by using 1-by-$N (\gg 1)$ optical router plus $N$ SPDs or a highly efficient photon-number resolving detector. The length of fibers of interfering paths are equalized within 1 cm to suppress the effect of dispersion, and the birefringence caused by fiber curvature is compensated by fiber paddle polarization controllers (FPCs) and combinations of quarter-, half-, and quarter-wave plates (QHQ). See Methods for the procedures to set the FPCs and the QHQs (denoted as PC and $\Theta_i$). Interference filters (IFs, half-maximum bandwidth of 3 nm at $s_{1,2}$ and $s_{1',2'}$ and 20 nm at $r_{1,2}$) at the end of each path in Fig.~\ref{fig:setup}(a) maintain the indistinguishability between photons from independent pairs.

\begin{figure*}[t]
\centering
\includegraphics[width=6 in]{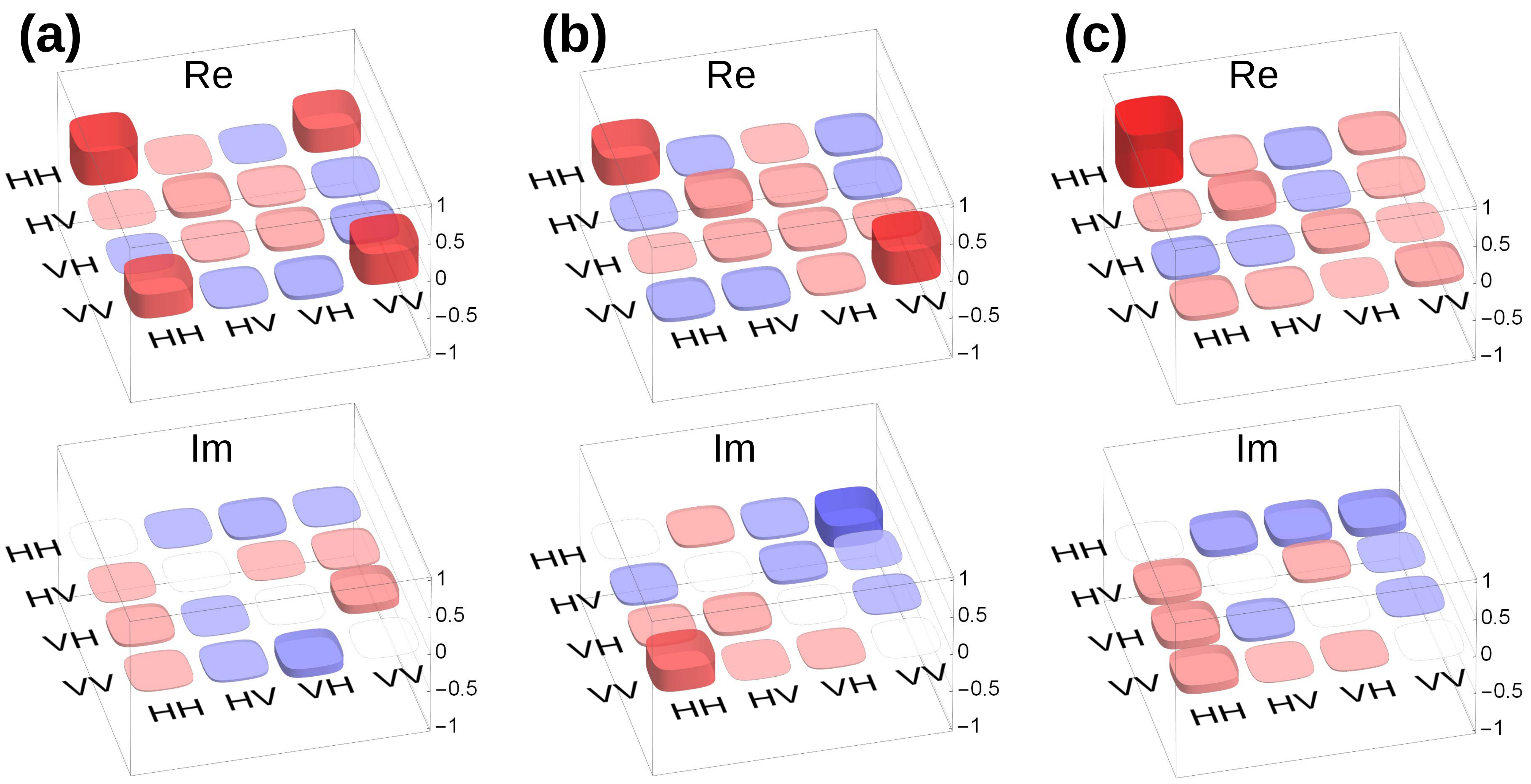}
\caption{Reconstructed density matrices of teleported output states: (a) $\ket{H}\ket{H}+\ket{V}\ket{V}$, (b) $\ket{H}\ket{H}+i\ket{V}\ket{V}$, and (c) $\ket{H}\ket{H}$.}
\label{fig:state}
\end{figure*}

The teleported two photons proceed to the receiver modes $r_1$ and $r_2$. Their polarization states are measured by fiber-based polarization analyzers shown in Fig.~\ref{fig:setup}(c).
% When characterizing the input(output) states in modes $s(r)_1$ and $s(r)_2$ or the four-photon GHZ state,
A QWP and a HWP (noted ``MB'') at each entrance of photons (on the left) in Fig.~\ref{fig:setup}(b) and \ref{fig:setup}(c) set the polarization basis measured by the SPDs when characterizing the initial state, the four-photon GHZ state, and the final state. During the teleportation experiments, the measurement bases of BSAs are set as $X$ to fix the detectable Bell sates as $(\ket{\psi^-}, \ket{\phi^-})$. Two-, four-, and six-fold coincidences of total 20 SPDs in Fig.~\ref{fig:setup}(a) are analyzed by an FPGA-based logic unit.

%{\it Input \& output states of teleportation}: 
We first generate three input states for teleportation, (a) $\ket{H}\ket{H}+\ket{V}\ket{V}$, (b) $\ket{H}\ket{H}+i\ket{V}\ket{V}$, (c) $\ket{H}\ket{H}$, and a four-photon GHZ state (See Supplementary Information for the measured initial states). Then we measure the teleported output states. Six-fold CCs are recorded while varying the measurement bases in modes $r_1$ and $r_2$ for QST. The unit counting period was 20 h for each basis and the average CC was 1.5 cph. To complete the teleportation protocol, unitary Pauli operations are applied to the received photons by re-arranging the count records among the QST measurement bases ($XX$, $XY$, \dots, $ZZ$), depending on the results of $\rm {\bf B}$'s. For example, when $\rm {\bf B}$'s result in one $\ket{\phi^-}$ and one failure event, the measured logical Bell state is $\ket{\Phi^-_L}$, therefore, to compensate for the local phase-flip operation, the count records measured on the $X$ and $Y$ bases in mode $r_1$ (or mode $r_2$) are exchanged between +1 and -1 data. The reconstructed three output states are shown in Figs.~\ref{fig:state}(a--c). The fidelities between the teleported states in Figs.~\ref{fig:state}(a--c) and the input states are (a) 0.84(4), (b) 0.78(6) and (c) 0.75(5), and exceed the classical bound (2/3) by 1.7 -- 4.3 standard deviations. The infidelities can be attributed to imperfections of the initial input and GHZ states and other experimental errors (See Supplementary Information).

%\section{Discussion}

Our protocol differs from the previous designs \cite{Sun16,Pant17} in which a trusted node plays a major role to connect the participants and transfers the information. It may be useful to establish a long distance quantum communication via distributed nodes, none of which necessarily relays the full quantum information. This is also applicable to the storage and retrieval of quantum secret with spatially separate quantum memories \cite{Choi2008}.
Our work can be extended further to be fault-tolerant by error correction encoding against photon losses, operation errors, and dishonest participants. For example, a parity state encoding \cite{Ralph05} can be employed to correct the effects of photon losses, errors and dishonest parties to some extent (See Supplementary Information). It allows, in principle, to transfer quantum information with arbitrarily high success probabilities even under losses and errors \cite{SLee18,Ewert16}. Verification strategies of multipartite entanglement \cite{Pappa12,McCutcheon16} are useful to prepare the entangled network in the presence of dishonest parties. Combination with other advanced protocols \cite{TQSS1,GQSS1} are further considerable. Finally, our Bell-state analysis would be applicable to other recently advanced quantum communication protocols \cite{SLee18,Ewert16,Azuma15}.

\section{Methods}

{\bf Distributed Bell-state measurement.}  Let us first consider the Bell states of the two-photon logical qubits, $\ket{\Phi^{\pm}_L}=\ket{0_L}\ket{0_L}\pm \ket{1_L}\ket{1_L}=\ket{H}_{1}\ket{H}_{2}\ket{H}_{1'}\ket{H}_{2'}\pm\ket{V}_{1}\ket{V}_{2}\ket{V}_{1'}\ket{V}_{2'}$ and $\ket{\Psi^{\pm}_L}=\ket{0_L}\ket{1_L} \pm \ket{1_L}\ket{0_L}=\ket{H}_{1}\ket{H}_{2}\ket{V}_{1'}\ket{V}_{2'}\pm\ket{V}_{1}\ket{V}_{2}\ket{H}_{1'}\ket{H}_{2'}$, where the subscript $j$ ($j'$) denotes the $j$-th photon in the first (second) qubit. They can be represented, by rearranging the modes, as $\ket{\Phi^{\pm}_L}=\ket{\phi^+}_{11'}\ket{\phi^{\pm}}_{22'}+\ket{\phi^-}_{11'}\ket{\phi^{\mp}}_{22'}$ and $\ket{\Psi^{\pm}_L}=\ket{\psi^+}_{11'}\ket{\psi^{\pm}}_{22'}+\ket{\psi^-}_{11'}\ket{\psi^{\mp}}_{22'}$, by two-photon Bell states $\ket{\phi^\pm}=\ket{H}\ket{H}\pm\ket{V}\ket{V}$ and $\ket{\psi^\pm}=\ket{H}\ket{V}\pm\ket{V}\ket{H}$ within $(1,1')$ and $(2,2')$ modes. Performing two independent $\rm {\bf B}$ on $(1,1')$ and $(2,2')$ modes (each can discriminate two Bell states out of four \cite{SBM1,SBM2}) identifies the logical Bell states. 

The logical Bell states with $n$-photon encoding are $\ket{\Phi^{\pm}_L}=\ket{H}_{1}\cdots\ket{H}_{n}\ket{H}_{1'}\cdots\ket{H}_{n'}\pm\ket{V}_{1}{\small \cdots}\ket{V}_{n}\ket{V}_{1'}{\small \cdots}\ket{V}_{n'}$ and $\ket{\Psi^{\pm}_L}=\ket{H}_{1}\cdots\ket{H}_{n}\ket{V}_{1'}\cdots\ket{V}_{n'}\pm\ket{V}_{1}\cdots\ket{V}_{n}\ket{H}_{1'}\cdots\ket{H}_{n'}$, where the first $n$ photons (from $1$ to $n$) carry one qubit and the following $n$ photons (from $1'$ to $n'$) carry the other qubit. Re-arrange the order of modes $(1,\ldots,n,1',\ldots,n')$ to $(1,1',2,2',\ldots,n,n')$ leads to
\begin{equation}
\begin{aligned}
\nonumber
&\ket{\Phi^{+(-)}_L}=\sum_{j={\rm even(odd)} \leq n}{\cal P}[\ket{\phi^-}^{\otimes j}\ket{\phi^+}^{\otimes n-j}],\\
&\ket{\Psi^{+(-)}_L}=\sum_{j={\rm even(odd)} \leq n}{\cal P}[\ket{\psi^-}^{\otimes j}\ket{\psi^+}^{\otimes n-j}],
\end{aligned}
\end{equation}
where ${\cal P}[\cdot]$ denotes the sum of all possible permutation e.g.~${\cal P}[\ket{\phi^-}\ket{\phi^+}^{\otimes 2}]=\ket{\phi^-}\ket{\phi^+}\ket{\phi^+}+\ket{\phi^+}\ket{\phi^-}\ket{\phi^+}+\ket{\phi^+}\ket{\phi^+}\ket{\phi^-}$. Based on the results of $n$-times of $\rm {\bf B}$  performed on modes $(j,j')$ ($j=1,...,n$), one can identify the logical Bell states. If the results of $\rm {\bf B}$, which can discriminate $\ket{\phi^-}$ and $\ket{\psi^-}$, include even (odd) number of $\ket{\phi^-}$, the logical Bell state is $\ket{\Phi^+_L}$ ($\ket{\Phi^-_L}$). If the results of $\rm {\bf B}$ include even (odd) number of $\ket{\psi^-}$, the logical Bell state is $\ket{\Psi^+_L}$ ($\ket{\Psi^-_L}$). It fails only when all performed $\rm {\bf B}$'s fail, therefore the success probability is $1-2^{-n}$. It indicates that increasing $n$ boosts the probability to successfully discriminate the logical Bell states \cite{SLee15}.

{\bf Accessible information.} The accessible information to a sender $s_j$, attempting to reconstruct the secret $\alpha\ket{H}+\beta\ket{V}$ at his/her location during the teleportation process based on all the other senders' results, can be obtained as follows. After all the other senders except $s_j$ perform $\rm {\bf B}$, the remaining state is either $\ket{\phi^-}_{s_j}(\alpha\ket{H}+\beta\ket{V})_r+\ket{\phi^+}_{s_j}(\alpha\ket{H}-\beta\ket{V})_r$ or $\ket{\psi^-}_{s_j}(\alpha\ket{V}+\beta\ket{H})_r+\ket{\psi^+}_{s_j}(\alpha\ket{V}-\beta\ket{H})_r$ (here $m=1$ for simplicity). By tracing out the receiver's party, the reduced state arrives at either $|\alpha|^2\ket{H,H}\bra{H,H}+|\beta|^2\ket{V,V}\bra{V,V}$ or $|\alpha|^2\ket{H,V}\bra{H,V}+|\beta|^2\ket{V,H}\bra{V,H}$. It shows that only the amplitude information of the secret is accessible to $s_j$.

{\bf Polarization-entangled photon-pair sources.}
Photons are generated by non-collinear degenerate type-I SPDC in a pair of two mutually orthogonal BBO crystals with a pulsed pump laser (average power 200 mW, repetition 76 MHz, center wavelength 390 nm, half-maximum bandwidth 2 nm). The pump polarization is set to be 45$^\circ$ before entering the first BBO crystal pair to equalize the generation rates of horizontal photon pairs and vertical photon pairs. To remove slight spatial and temporal mismatch between the two photon pairs, spatial compensators (SCs, not shown in Fig.~\ref{fig:setup}(a)) and temporal compensators (TCs) made of BBO and quartz crystals, respectively, are added according to the design in \cite{Rangarajan09}. Walk-off compensators (WCs) in Fig.~\ref{fig:setup}(a) compensate for the polarization-dependent spatial walk-off of the pump beam caused by the SPDC crystals.

{\bf SMF-based BSA.}
The optical-fiber-based BSA in Fig.~\ref{fig:setup}(b) has the same basic structure (1 BS + 2 PBS) as the original BSA~\cite{Mattle96} with additional BSs and SPDs to identify the failure events. Unpredictable amount of birefringence of the optical fiber sections are compensated for by adjusting polarization controllers (PCs) made of either a QHQ set or a FPC as shown in Fig.~\ref{fig:setup}(b). Instead of using interference of two polarization-modulated coherent light sources~\cite{Jennewein02} or a source of polarization singlet state~\cite{SMLee18}, we use a step-wise method to first fix the principal polarizations and secondly adjust the birefringence between them as follows: (i) Adjust the PC and two FPCs such that $H$-($V$-)polarized light entering modes $s_{i}$ and $s_{i'}$ exits through output ports $h_j$ ($v_j$). (ii) Inject light backward through mode $h_1$, and, by slightly misaligning the FPC from the optimal position, make the relative magnitudes of $H$ and $V$ polarizations at modes $s_i$ and $s_{i'}$ be similar. (iii) Rotate the HWP between the two QWPs whose slow axes are along 45$^\circ$ in $\Theta_i$, based on polarimetry, to equalize two relative phases between $H$ and $V$ polarizations of the backward propagating light at $s_i$ and $s_{i'}$. (iv) Return the FPC to the optimal position. The MBs determine the detectable Bell states. For example, by changing the measurement bases of the SPDs from $Z$ to $X$, the successfully measured Bell states becomes $\ket{\psi^-}$ and $\ket{\phi^-}$ as in our case, in contrast to $\ket{\psi^-}$ and $\ket{\psi^+}$ in the conventional BSA.

%%%%%%%%%%%%%%%%%%%%%%%%%%%%%%%%%%%%%%%%%%%%%%%%%%%%%%%%%%%%\
%\section*{References}

\section{Acknowledgments}

S.M.L. and H.S.P. acknowledge the support of the R\&D Convergence program of NST of Republic of Korea (CAP-18-08-KRISS) and the KRISS project (KRISS-2018-GP2018-0017). H.J. was supported by a National Research Foundation of Korea grant funded by the Korea government (Grant No. 2010-0018295) and by the Korea Institute of Science and Technology Institutional Program (Project No. 2E26680-16-P025).

\section{Author Contributions}

S.-W.L., H.J., and H.S.P. conceived the work. S.M.L. designed and carried out the experiment. S.-W.L. theoretically completed the protocols. S.M.L. and S.-W.L. wrote the first draft. All the authors discussed the results and contributed to the editing of the manuscript.

\section{Competing Interests} The authors declare that they have no competing financial interests.

\section{Correspondence} 
Correspondence and requests for materials should be addressed to S.-W.L (email:swleego@gmail.com) or H.J (email:h.jeong37@gmail.com) or H.S.P (email:hspark@kriss.re.kr).

%%%%%%%%%%%%%%%%%%%%%%%%%%%%%%%%%%%%%%%%%%%%%%%%%%%%%%%%%%%%%%%%%%%%%%%%%%%%%%%%%%%%%%%%%%%%%%%%%%%%%%%%%%%%%%%%%%%%%%%%%%%%%%%%%%%%%%%%%%%%%%%%%%%%%%%%%%%%%%%%%%%%%%%%%%%%%%%%%%%%
%%%%%%%%%% Supplemental materials %%%%%%%%%%%%%%%%%%%%%%%%%%%%%%%%%%%%%%

\pagebreak
\newpage
\newpage

\widetext
%\par
\begin{center}
\textbf{\large Supplementary Material}
\end{center}
%%%%%%%%%% Merge with supplemental materials %%%%%%%%%%
%%%%%%%%%% Prefix a "S" to all equations, figures, tables and reset the counter %%%%%%%%%%
\setcounter{equation}{0}
\setcounter{figure}{0}
\setcounter{table}{0}
\setcounter{section}{0}
%\setcounter{page}{1}
%\makeatletter
\renewcommand{\theequation}{S\arabic{equation}}
\renewcommand{\thefigure}{S\arabic{figure}}
\renewcommand{\thetable}{S\arabic{table}}
\renewcommand{\bibnumfmt}[1]{[S#1]}
\renewcommand{\citenumfont}[1]{S#1}

\section{Reconstructed four-photon GHZ state}
\begin{figure}[h]
\centering
\includegraphics[width=5.5 in]{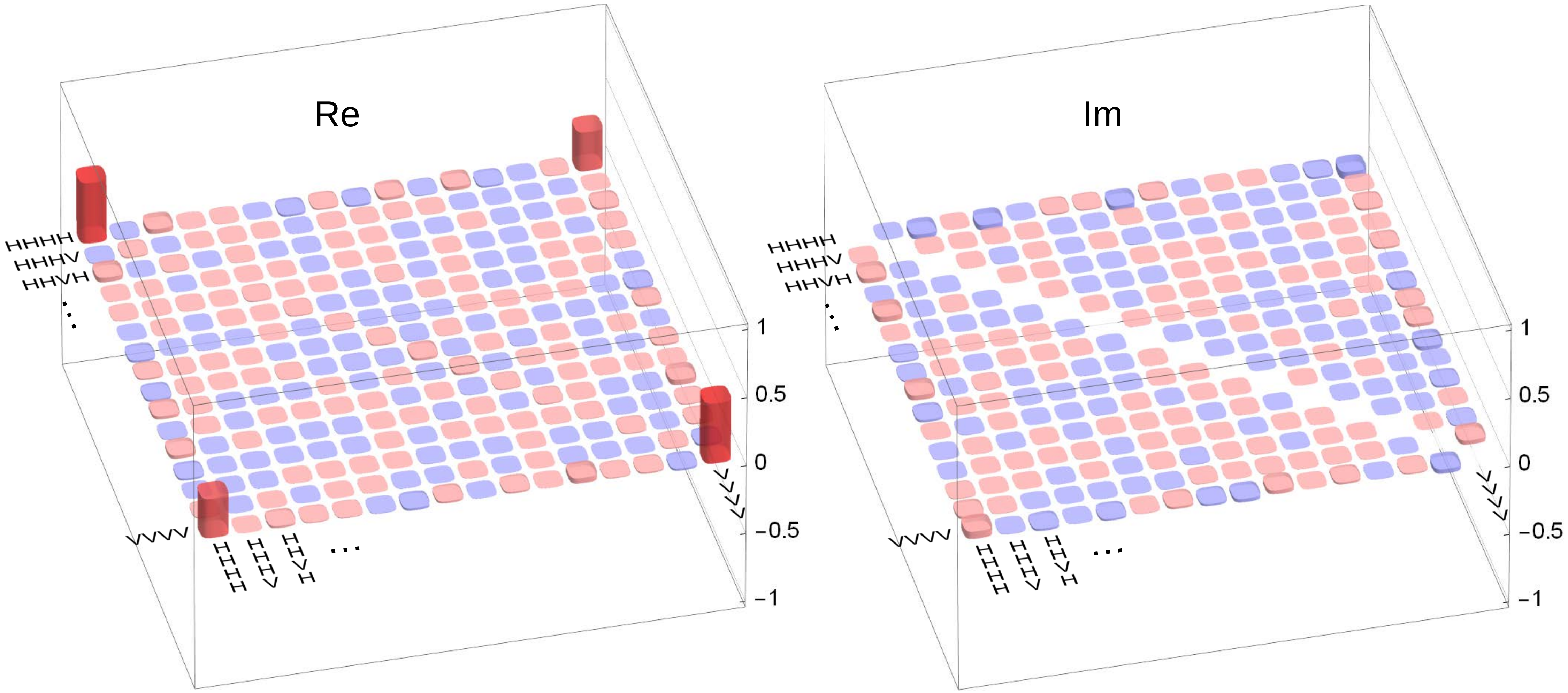}
\caption{Reconstructed density matrice for the four-photon GHZ state, $\ket{H}^{\otimes 4}+\ket{V}^{\otimes 4}$.}
\label{fig:GHZ}
\end{figure}

Figure~\ref{fig:GHZ} shows the reconstructed density matrix of the four-photon GHZ state. Quantum state tomography (QST) with maximum likelihood estimation~\cite{James01} was performed based on $3^4 = 81$ measurement basis sets with a unit acquisition time of 100 s. Four-fold CCs for the $\ket{H}^{\otimes 4}$ or $\ket{V}^{\otimes 4}$ components were 1.5 cps. Fidelity $F = \bra{\textrm{GHZ}_4} \hat{\rho} \ket{\textrm{GHZ}_4}$ between the reconstructed state $\hat{\rho}$ and the ideal state $\ket{\textrm{GHZ}_4}$ was 0.73(1). 

\begin{figure}[h]
\centering
\includegraphics[width=5.5 in]{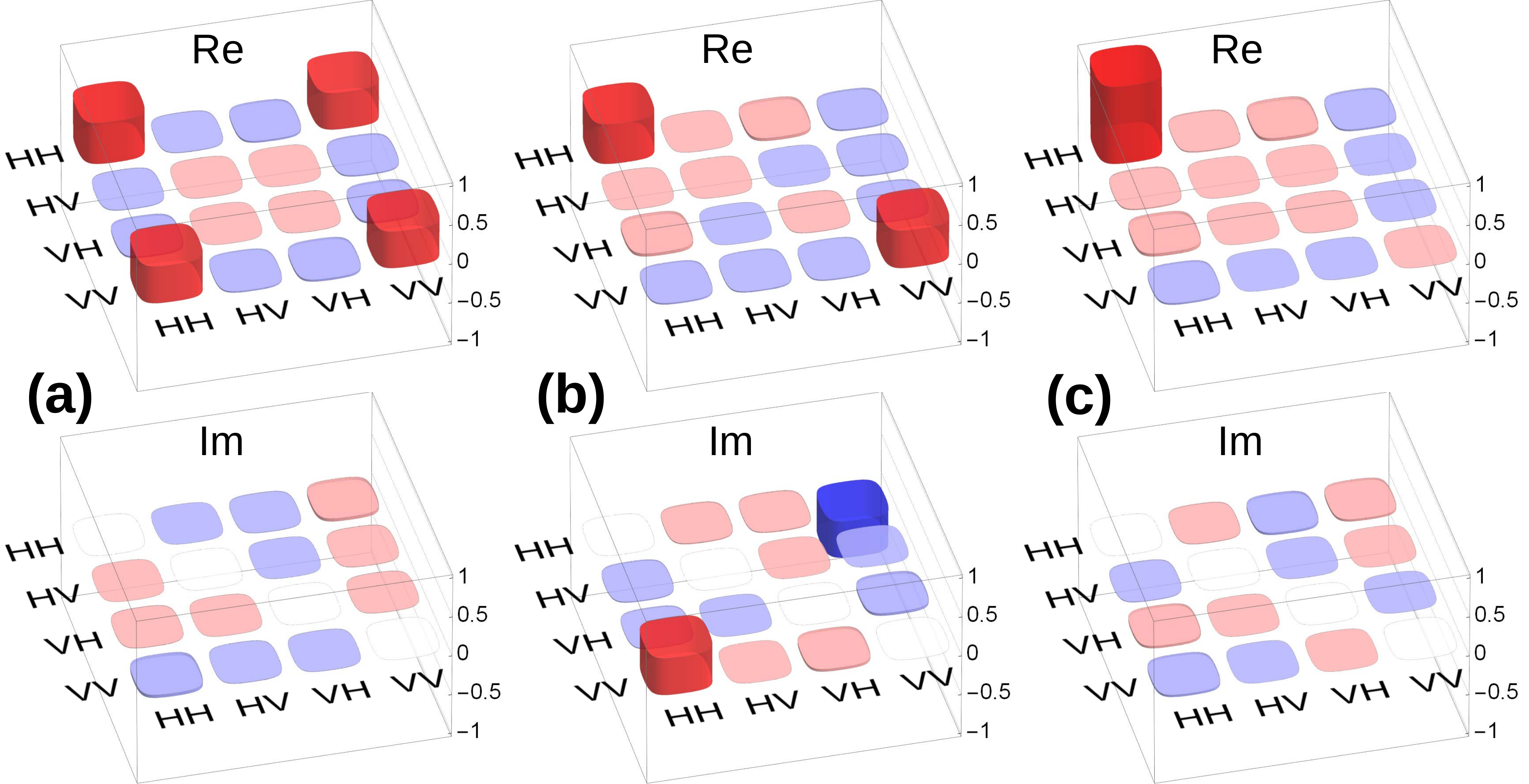}
\caption{Reconstructed density matrices of three input states for (a) $\ket{HH}+\ket{VV}$, (b) $\ket{HH}+i\ket{VV}$, and (c) $\ket{HH}$.}
\label{input}
\end{figure}

\section{Reconstructed input states \& fidelity bounds}

Figure~\ref{input} shows the reconstructed density matrices of three input states (a) $\ket{HH}+\ket{VV}$, (b) $\ket{HH}+i\ket{VV}$, and (c) $\ket{HH}$. These initial states were characterized by two-photon QST utilizing the SPDs in the two BSAs. The average two-fold CCs for the $\ket{H}^{\otimes 2}$ or $\ket{V}^{\otimes 2}$ components of the input states were 7 kcps. The results show fidelities with the ideal states of (a) 0.969(1), (b) 0.967(1), and (c) 0.992(1), respectively, as summarized in row $(\alpha)$ of Table \ref{fidtab}. Row $(\beta)$ of Table \ref{fidtab} shows our main results, fidelities between the experimentally reconstructed input and output states, which range from 0.75(5) to 0.84(4). These fidelities between the transferred states are higher than the fidelity between the experimentally reconstructed four-photon GHZ state and its ideal state, 0.73(1). To verify that this is a reasonable result and to assess the quality of the teleportation processes, we calculate the output states from the experimental input states and the GHZ state under the assumption of ideal Bell-state measurements. The calculated states are compared with the experimental input states to yield fidelities shown in row $(\gamma)$ of Table \ref{fidtab}. The calculated fidelities clearly exceed the fidelity of the GHZ state, and slightly higher than the values in row $(\beta)$. This reduction (0.01 -- 0.19) of fidelities from row $(\gamma)$ to row $(\beta)$ manifests the imperfections of our BSAs and unitary operations. Thus our experimental results of quantum teleportation of the entangled states reasonably agree with the theoretical expectations considering experimental imperfections, show the fidelities exceeding the classical bound (2/3) with at least 1.7 standard deviations without subtracting backgrounds.

In addition, we calculate fidelities between the ideal inputs and the expected output states based on the experimentally reconstructed GHZ state, which depend on the input states and range 0.762--0.927 as shown in row $(\delta)$ of Table \ref{fidtab}. These values are similar with (slightly (0.03--0.04) higher than) the fidelity of the GHZ state with the ideal state except for the case of (c) without superposition, and less than the row $(\gamma)$ for the mixed inputs. Finally, the row $(\epsilon)$ shows the fidelities between the ideal inputs and the experimentally reconstructed outputs.

\begin{table}[h]
\begin{tabular}{|c c|c|c|c|}
 \hline			& Fidelity between 							&~~case (a)~~&~~case (b)~~&~~case (c)~~\\\hline
($\alpha$)		& ${\rm in}_{exp}$, ${\rm in}_{ideal}	$ 	& 0.969(1) 	& 0.967(1) 	& 0.992(1)	\\\hline
($\beta$)		& ${\rm in}_{exp}$, ${\rm out}_{exp}	$	& 0.84(4) 	& 0.78(6) 	& 0.75(5)	\\\hline
($\gamma$) 	& ${\rm in}_{exp}$, ${\rm out}_{cal(ie)}	$	& 0.856 		& 0.868 		& 0.945 		\\\hline
($\delta$) 		& ${\rm in}_{ideal}$, ${\rm out}_{cal(ii)}	$	& 0.762 		& 0.768 		& 0.927 	\\\hline
($\epsilon$) 		& ${\rm in}_{ideal}$, ${\rm out}_{exp}	$	& 0.75(5) 		& 0.69(7) 		& 0.71(6)
\\\hline
\end{tabular}
\caption{Fidelities among experimentally measured, ideal and calculated states depending on considered states of (a)
$\ket{HH}+\ket{VV}$, (b) $\ket{HH}+i\ket{VV}$, and (c) $\ket{HH}$. $ie(ii)$ stands for state of ${\rm
in}_{exp(ideal)}$.}
\label{fidtab}
\end{table}

\section{Experimental errors}

The success probability of the teleportation can ideally reach 75\%~\cite{SLee15}, considering the three cases occurring 25\%: success/success (SS), success/failure (SF), and failure/success (FS) for the two $\rm {\bf B}$'s. However, since the probability to detect failure of each $\rm {\bf B}$ is only 50\% with the current setup using eight SPDs, and the overall success probability decreases to 50\%. The six-fold coincidences recorded during the teleportation experiments can be categorized as five cases: SS, SF, FS, failure/failure (FF), and error (E) counts.
%that appear, for example, one $\rm B_s$ results in $\ket{\Phi^-}$ and the other $\rm B_s$ results in $\ket{\Psi^-}$.
If the optical components are ideal and there are no redundant photons, the relative probabilities of the five cases are 4/9, 2/9, 2/9, 1/9, and 0, respectively. Averaged over the three teleported output states, these relative frequencies (standard deviation) were 36.8(1.6)\%, 19.4(1.3)\%, 16.6(1.3)\%, 5.1(0.7)\%, and 22.1(1.4)\%. The relative magnitudes for only SS, SF, FS, and FF cases are 47.3(1.9)\%, 24.9(1.7)\%, 21.3(1.6)\%, and 6.5(1)\%, respectively, and agree reasonably with the theoretical expectation (4:2:2:1). We expect that the error probability is due to the contribution of higher-order photon pair generation by SPDC and imperfect extinction ratios ($\geq 20$ dB) of the polarization controlling optical components. The fidelities of the \textit{rescued} teleported states only collecting the CCs with one of the $\rm {\bf B}$'s failing were (a) 0.80(6), (b) 0.67(9) and (c) 0.65(7).

\section{Fault-tolerant teleportation of shared quantum secret}

\subsection{Decomposition of the encoded Bell states}

We extend the proposed teleportation protocol between multiple parties in quantum network to include an error correction code. As photon losses are particularly a major detrimental factor in photonic quantum communication, we employ a parity state encoding \cite{Ralph05}, instead of a GHZ state encoding, to protect qubits from the loss-induced decoherence. In this encoding, the logical bases are given as $\ket{0_L}=\ket{+^{(p)}}^{\otimes n}$ and $\ket{1_L}=\ket{-^{(p)}}^{\otimes n}$, where $\ket{\pm^{(p)}}=\ket{+}^{\otimes p}\pm\ket{-}^{\otimes p}$ and $\ket{\pm}=\ket{H}\pm\ket{V}$. Each logical qubit contains $n$ blocks of $p$ photons (total $N=np$).
The logical Bell states $\ket{\Phi^{\pm}_L}=\ket{0_L}\ket{0_L}\pm\ket{1_L}\ket{1_L}$ and $\ket{\Psi^{\pm}_L}=\ket{0_L}\ket{1_L}\pm\ket{1_L}\ket{0_L}$, can then be completely decomposed into the block-size Bell states, $\ket{\phi^{\pm}_{(p)}}=\ket{+^{(p)}}\ket{+^{(p)}}\pm\ket{-^{(p)}}\ket{-^{(p)}}$ and $\ket{\psi^{\pm}_{(p)}}=\ket{+^{(p)}}\ket{-^{(p)}}\pm\ket{-^{(p)}}\ket{+^{(p)}}$ (see Ref.~\cite{SLee18} for the details). Likewise, each block-size Bell state can also be completely decomposed into the two-photon Bell states, $\ket{\phi^{\pm}}=\ket{+}\ket{+}\pm\ket{-}\ket{-}$ and $\ket{\psi^{\pm}}=\ket{+}\ket{-}\pm\ket{-}\ket{+}$. Let us refer to the logical, block-size, photon-wize Bell states as the 2nd-, 1st-, 0th-level Bell states, respectively. A Bell state in a higher level can be fully characterized by the type and number of lower level Bell states in its decomposition (see Table~\ref{table:BMresult}).

%%%%%%%%%%%%%%%%%%%%%%%%%%%%%%%%%%%%%%%%%%%%
\begin{table}[h]
\caption{\label{table:BMresult} \textbf{Bell states decomposition}}
\centering
%\begin{ruledtabular}
\begin{tabular}{|l  |c |  c|}
%\hline
\hline
Level & Bell states  & Decomposed into\\[0.5ex]
\hline
2nd  & & {\small even(odd) number of} $\ket{\phi^{-}_{(p)}}$, \\[-2ex]
(logical) & \raisebox{1.5ex}{$\ket{\Phi^{+(-)}_L}$} &and $\ket{\phi^{+}_{(p)}}$ for others\\
\cline{2-3}
& & even(odd) number of $\ket{\psi^{-}_{(m)}}$,\\[-2ex]
& \raisebox{1.5ex}{$\ket{\Psi^{+(-)}_L}$} &and $\ket{\psi^{+}_{(p)}}$ for others\\
\hline
1st &  & even number of $\ket{\psi^{+(-)}}$, \\[-2ex]
(block) & \raisebox{1.5ex}{$\ket{\phi^{+(-)}_{(p)}}$} &and $\ket{\phi^{+(-)}}$ for others\\
\cline{2-3}
 & & odd number of $\ket{\psi^{+(-)}}$, \\[-2ex]
 & \raisebox{1.5ex}{$\ket{\psi^{+(-)}_{(p)}}$} &and $\ket{\phi^{+(-)}}$ for others\\[-0.5ex]
\hline
\end{tabular}
%\end{ruledtabular}
\end{table}

%%%%%%%%%%%%%%%%%%%%%%%%%%%%%%%%%%%%%%%%%%%%%
\subsection{Concatenated Bell-state measurement}

%%%%%%%%%%%%%%%%%%%%%%%%%%%%
\begin{figure}[b]
\centering
\includegraphics[width=5 in]{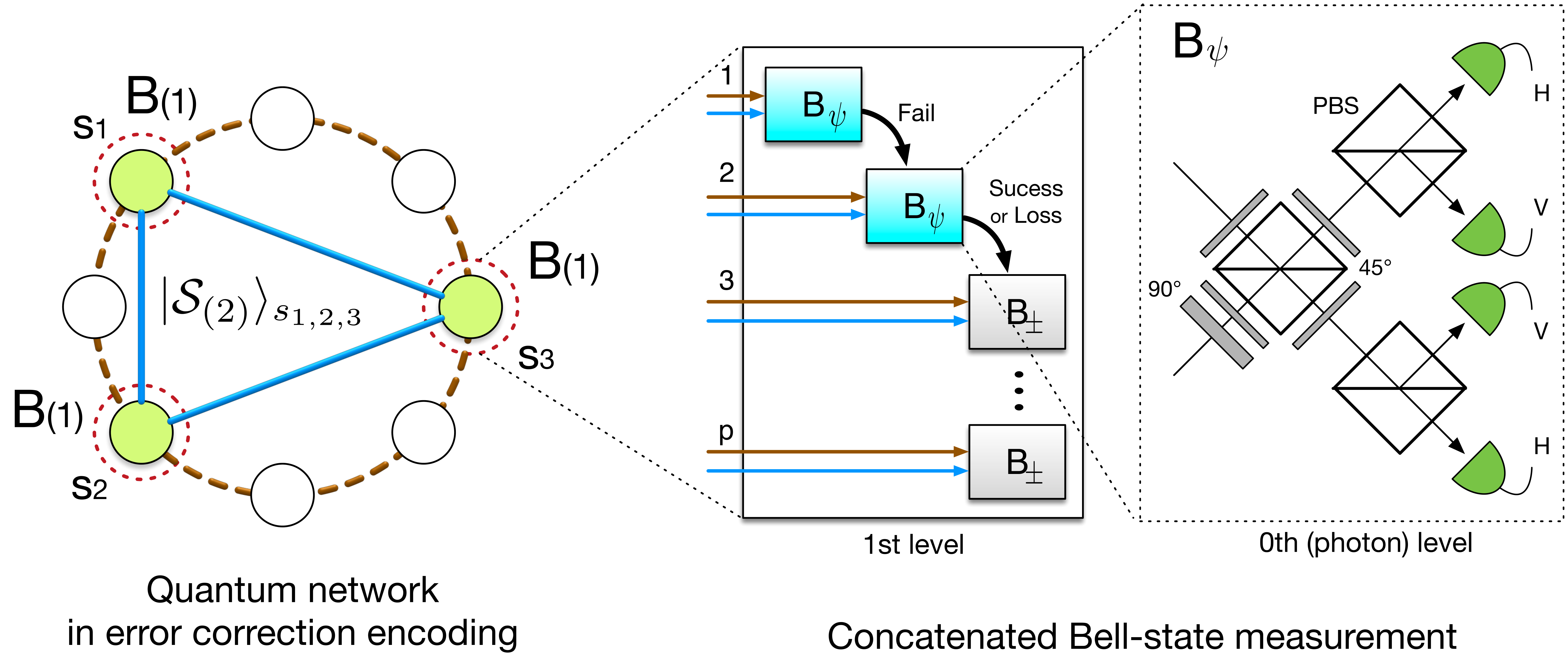}
\caption{Fault-tolerant quantum teleportation of shared quantum secret between multiple parties in quantum network with parity-state error correction encoding. Each sender sharing a quantum secret $\ket{{\cal S}_{(2)}}_s$ performs a logical Bell-state measurement, ${\rm {\bf B}}_{(1)}$, composed of $p$ times of standard versions of Bell-state measurements ${\rm {\bf B}}_{(0)}$, in a concatenated manner, and announces the result. There are three different types of ${\rm {\bf B}}_{(0)}$, i.e.,~$\{{\rm {\bf B}}_{\psi}, {\rm {\bf B}}_{+}, {\rm {\bf B}}_{-}\}$ discriminating $\{(\ket{\psi^+},\ket{\psi^-}), (\ket{\phi^+},\ket{\psi^+}), (\ket{\phi^-},\ket{\psi^-})\}$ respectively. Receivers can reconstruct and share the secret by a logical Pauli flip performed through a joint work of local operations. This protocol can succeed with arbitrarily high probabilities even with photon losses, errors and dishonest participants.}
\label{fig:NNTLT}
\end{figure}
%%%%%%%%%%%%%%%%%%%%%%%%%%%%

We now introduce an advanced Bell-state measurement scheme referred to as concatenated Bell-state measurement (CBM) \cite{SLee18}. This is composed of three levels of linear-optic Bell-state measurements as illustrated in Fig.~\ref{fig:NNTLT}, where the 0th, 1st, and 2nd levels comprise photon-size, block-size, and logically encoded qubits, respectively:

\textbf{[\textit{0th level}]} The 0th level, ${\rm {\bf B}}_{(0)}$, employs the standard scheme of Bell-state measurement using linear optics elements such as beam splitters, polarizing beam splitters, wave plates and single-photon detectors \cite{SBM1,SBM2}. It enables to unambiguously discriminate two out of the four two-photon Bell states, $\ket{\phi^{\pm}}$ and $\ket{\psi^{\pm}}$. The two identified Bell states can be chosen by modifying the wave plates at the input modes. We define three different types $\{{\rm {\bf B}}_{\psi}, {\rm {\bf B}}_{+}, {\rm {\bf B}}_{-}\}$ that respectively discriminate $\{(\ket{\psi^+},\ket{\psi^-}), (\ket{\phi^+},\ket{\psi^+}), (\ket{\phi^-},\ket{\psi^-})\}$.

\textbf{[\textit{1st level}]} In the 1st level, ${\rm {\bf B}}_{(1)}$, total $p$ times of ${\rm {\bf B}}_{(0)}$ are performed. It starts with ${\rm {\bf B}}_{\psi}$ on one photon pair (one from the first qubit and the other from the second qubit). ${\rm {\bf B}}_{\psi}$ is applied repeatedly until it succeeds or detects a loss or consecutively fails $q$-times ($0\leq q\leq p-1$). Either ${\rm {\bf B}}_{+}$ or ${\rm {\bf B}}_{-}$ is applied on the remaining photon pairs; ${\rm {\bf B}}_{\pm}$ is applied if ${\rm {\bf B}}_{\psi}$ succeeded with $\ket{\psi^\pm}$, while arbitrary one (either ${\rm {\bf B}}_{+}$ or ${\rm {\bf B}}_{-}$) is chosen if loss is detected or $q$-time failures happen. Note that $q$ can be optimized for a given $N$ and loss rate $\eta$. ${\rm {\bf B}}_{(1)}$ leads to the following three results: (a) {\it Success} - Full discrimination of $\ket{\phi^{\pm}_{(p)}}$ and $\ket{\psi^{\pm}_{(p)}}$ is possible unless loss occurs. For example, suppose that a ${\rm {\bf B}}_{\psi}$ succeeds with $\ket{\psi^{+}}$. Then, subsequently performed ${\rm {\bf B}}_{+}$ on the remaining pairs should yield either $\ket{\phi^{+}}$ or $\ket{\psi^{+}}$. By counting all the results, if even(odd) number of $\ket{\psi^{+}}$ appear, we can find that the 1st level Bell state is $\ket{\phi^{+}_{(p)}}$($\ket{\psi^{+}_{(p)}}$) from the Table~\ref{table:BMresult}. (b) {\it Sign $\pm$ discrimination} - We note that the sign $\pm$ of the 1st level Bell state can be identified as long as any single $\rm {\bf B}_{\psi}$ succeeds or any one $\rm {\bf B}_{\pm}$ is performed without loss. (c) {\it Failure} - It fails when no $\rm {\bf B}_{\psi}$ succeeds and loss occurs in all performed ${\rm {\bf B}}_{\pm}$.

\textbf{[\textit{2nd level}]} The logical level, $\rm {\bf B}_{(2)}$, is composed of $n$ times of $\rm {\bf B}_{(1)}$, which can be performed independently in separated locations. The logical Bell states, $\ket{\Phi^{\pm}_L}$ and $\ket{\Psi^{\pm}_L}$, can be unambiguously identified as long as any single $\rm {\bf B}_{(1)}$ succeeds and no $\rm {\bf B}_{(1)}$ fails. The result of the logical Bell-state measurement is given by $\ket{\{\Phi,\Psi\}_L^{(-)^s}}$, in which the symbol $\Phi$ and $\Psi$ is discriminated by the result of any succeeded $\rm {\bf B}_{(1)}$, and the sign $(-)^s$ is identified if $s$ (either even or odd) number of minus($-$) signs appear among all $\rm {\bf B}_{(1)}$ results. For example, given the results of $\rm {\bf B}_{(1)}$ as $\{\ket{\phi_{(m)}^{-}},+,-\}$ when $n=3$, $\ket{\Phi^{+}}_L$ can be identified from the appeared symbol $\phi$ and even number of minus ($-$) sign.

\subsection{Fault-tolerant teleportation in quantum network}

We describe a fault-tolerant version of the proposed teleportation protocol between multiple parties. The scheme is illustrated in Fig.~\ref{fig:NNTLT}. We here employ entangled photons in the parity state encoding with $(n,p)$ as the channel of the quantum network instead of the GHZ-entanglement of photons. We assume that a sender group with arbitrary $n$ participants share a secret $\ket{{\cal S}_{(2)}}_s$ in the parity state encoding denoted by the subscript $(2)$. Each sender $s_j$ performs the 1st level Bell-state measurement $\rm {\bf B}_{(1)}$ on the photon pairs (one photon is from the secret and the other from the channel), and announce the result, either {\it Success} or {\it Sign $\pm$ discrimination} or {\it Failure}. Conditioned on the results of all the sender's $\rm {\bf B}_{(1)}$, the result of the logical Bell-state measurement can be identified from the Table.~\ref{table:BMresult}. According to the identified results, receivers can reconstruct the secret at their locations by an appropriate joint work of local operations. Remarkably, even if photons are lost in the secret or network channel and only one photon survives per qubit at each senders' location, $\rm {\bf B}_{(1)}$ can discriminate whether their sign is $(+)$ or $(-)$. Therefore, under the condition that at least one sender executes $\rm {\bf B}_{(1)}$ without loss, the logical Bell states can be unambiguously discriminated based on the results of all the members of the sender group.

In a realistic situation, depolarizing errors or imperfections in operation can produce logical errors (bit or sign flips) in Bell-state measurements. In addition, some senders may be dishonest and announce their results differently with the real results obtained by $\rm {\bf B}_{(1)}$. These lead to either sign ($+ \leftrightarrow -$) or symbol ($\phi \leftrightarrow \psi$) flip in the final result of the logical Bell-state measurement. We note that sign flip errors that occur in any $\rm {\bf B}_{(0)}$ can be corrected by majority vote, based on the fact that the signs in all $\rm {\bf B}_{(0)}$ results should be the same for an ideally performed $\rm {\bf B}_{(1)}$. Symbol flips due to errors or fraud can also be corrected to some extent eventually in the logical level from the fact the symbols of all $\rm {\bf B}_{(1)}$ in the result of ideally performed CBM are the same. This is because the parity state encoding is a generalized form of the Shor error correcting code \cite{Shor95}. Details of the performance of the parity state encoding and CBM in realistic situations are found in Ref.~\cite{SLee18}.

\end{document}